\documentclass[twocolumn, twoside, 10pt]{IEEEtran}
\usepackage{amsfonts}
\usepackage{mathrsfs}
\usepackage{amssymb,amsmath}
\usepackage{balance}

\usepackage{algorithm}
\usepackage{algorithmic}
\usepackage{amsmath,amssymb,epsfig,subfigure}
\usepackage{graphicx}
\usepackage{epstopdf}
\usepackage{theorem}
\usepackage{array,color}
\usepackage[compress]{cite}
\usepackage{enumerate}
\usepackage{bm}
\usepackage{balance}
\usepackage[symbol]{footmisc}

\theoremheaderfont{\normalfont\bfseries}

\begin{document}

\title{Deep Learning-Based Decoding of Constrained Sequence Codes}
\author{Congzhe~Cao, Duanshun Li,
        and~Ivan~Fair,~\IEEEmembership{Member,~IEEE}
\thanks{Manuscript received on 15 December 2018, revised on 5 April 2019. This work was supported by the Natural Science and Engineering Research
	Council (NSERC) of Canada and Alberta Innovates Technology
	Futures (AITF). The material in this paper was presented in part at the 2018 IEEE Globecom Workshops (GC Wkshps), Abu Dhabi, United Arab Emirates, 9-13 December 2018. \emph{(Corresponding author:
		Congzhe Cao.)}} 
\thanks{Congzhe Cao and Ivan Fair are with the Department
of Electrical and Computer Engineering, University of Alberta, Edmonton, AB T6G 1H9, Canada (email:\{congzhe, ivan.fair\}@ualberta.ca).
Duanshun Li is with the Department of Civil and Environmental Engineering, University of Alberta, Edmonton, AB T6G 1H9, Canada (email:\{duanshun\}@ualberta.ca)
}
}

\maketitle

\begin{abstract}
Constrained sequence (CS) codes, including fixed-length CS codes and variable-length CS codes, have been widely used in modern wireless communication and data storage systems. Sequences encoded with constrained sequence codes satisfy constraints imposed by the physical channel to enable efficient and reliable transmission of coded symbols. In this paper, we propose using deep learning approaches to decode fixed-length and variable-length CS codes. Traditional encoding and decoding of fixed-length CS codes rely on look-up tables (LUTs), which is prone to errors that occur during transmission. We introduce fixed-length constrained sequence decoding based on multiple layer perception (MLP) networks and convolutional neural networks (CNNs), and demonstrate that we are able to achieve low bit error rates that are close to maximum a posteriori probability (MAP) decoding as well as improve the system throughput. Further, implementation of capacity-achieving fixed-length codes, where the complexity is prohibitively high with LUT decoding, becomes practical with deep learning-based decoding. We then consider CNN-aided decoding of variable-length CS codes. Different from conventional decoding where the received sequence is processed bit-by-bit, we propose using CNNs to perform one-shot batch-processing of variable-length CS codes such that an entire batch is decoded at once, which improves the system throughput. Moreover, since the CNNs can exploit global information with batch-processing instead of only making use of local information as in conventional bit-by-bit processing, the error rates can be reduced. We present simulation results that show excellent performance with both fixed-length and variable-length CS codes that are used in the frontiers of wireless communication systems.

\end{abstract}

\section{Introduction}
Constrained sequence (CS) codes have been widely used to improve the performance and reliability of communication and data storage systems such as visible light communications, wireless energy harvesting, optical and magnetic recording systems, solid state drives, and DNA-based storage \cite{Textbook, energyHarvesting, VLC1,DNAStorageImmink}. Since the initial study of CS coding in Shannon's 1948 paper \cite{Shannon}, researchers have continuously sought to design efficient CS codes that achieve code rates close to capacity with low implementation complexity \cite{Textbook, Shannon,Franaszek, 8B10B_Ref,DC-freeRLL_Immink,FixedLength2,Motwani, BalancedModulation,FixedLength3,Pearson1,Pearson2,DNAStorageImmink,myDeepLearning,ImminkVLCodes,SystematicPearsonCoding,WeberBalancedCode,MyMinimalSet,My_MinimalSet,MyJSAC,myPearson,MyMultiState}. Most CS codes in the literature are fixed-length {(FL)} codes \cite{Franaszek, 8B10B_Ref,DC-freeRLL_Immink,FixedLength2,Motwani, BalancedModulation,FixedLength3,Pearson1,Pearson2,DNAStorageImmink,myDeepLearning}, while recent work shows that variable-length {(VL)} CS codes have the potential to achieve higher code rates with much simpler codebooks \cite{ImminkVLCodes,SystematicPearsonCoding,WeberBalancedCode,MyMinimalSet,My_MinimalSet,MyJSAC,myPearson,MyMultiState}.

{Look-up tables (LUTs)} are widely used for encoding and decoding {FL} CS codes that map length-$k$ source words to length-$n$ codewords. Although many good {FL} codes have been proposed and used in practical systems, {FL} CS codes often suffer from the following drawbacks: i) advantage usually is not taken of whatever error control capability may be inherent in CS codes, therefore they are prone to errors that occur during transmission; ii) the capacity of most constraints is irrational, therefore it is difficult to construct a CS codebook with rate $k/n$ that is close to capacity without using very large values of $k$ and $n$. However, with large $k$ and $n$ values, the time and implementation complexity of {LUTs} become prohibitive since a total of $2^k$ codewords exist in the codebooks of binary CS codes. Therefore, the design of practical capacity-achieving {FL} CS codes has been a challenge for many years.

In contrast, {VL} codes have the flexibility to map {VL} source words to {VL} codewords in the codebook, and are therefore better able to achieve capacity-approaching code rates with small $\bar{k}$ and $\bar{n}$ values, where $\bar{k}$ and $\bar{n}$ are the average length of source words and codewords in the codebook, respectively. Since {VL} CS codes can be designed as instantaneous codes \cite{MyMinimalSet,My_MinimalSet,MyJSAC,myPearson,MyMultiState}, conventional decoders perform codeword segmentation of the received sequence bit-by-bit by checking whether the sequence being processed is a valid codeword upon reception of each incoming bit. Few attempts have been made to \emph{batch-process} the received sequence by a {VL} CS decoder such that the entire sequence is segmented at once in the manner of \emph{one-shot} decoding. It is desirable to perform one-shot batch-processing within {VL} CS decoders since that approach will greatly improve the system throughput. Furthermore, given information of the entire batch instead of bit-by-bit decisions, the decoder is more likely to correct errors in the received sequence and maintain synchronization.

With the advancement of greater computational power and increasingly sophisticated algorithms, reinforcement learning (RL) has demonstrated impressive performance on tasks such as playing video games \cite{RL1} and Go \cite{RL2}. RL commonly uses Q-learning for policy updates in order to obtain an optimal policy that maps the state space to the action space \cite{sutton}, however, obtaining the update rule from {LUTs}, as traditionally has been done, becomes impossible with large state-action space. The invention of deep Q-networks that use deep neural networks (DNNs) to approximate the Q-function enables sophisticated mapping between the input and output, with great success \cite{deepRL}. Motivated by this approach, we hypothesized that it would be promising to replace look-up tables in both {FL} and {VL} CS codes with DNNs. Therefore, we propose using DNNs for {FL} and {VL} CS decoding to overcome the drawbacks outlined above.

Recently several works have reported the application of DNNs to the decoding of error control codes (ECCs) \cite{deeplearningECC1, deeplearningECC2, deeplearningECC3, deeplearningECC4, deeplearningECC5,deepCode}. A DNN enables low-latency decoding since it enables \emph{one-shot} decoding, where the DNN finds its estimate by passing	each layer only once \cite{deeplearningECC1, deeplearningECC3, deeplearningECC4}. In addition, DNNs can efficiently execute in parallel and be implemented with low-precision data types on a graphical processing unit (GPU), field programmable gate array (FPGA), or application specific integrated circuit (ASIC) \cite{deeplearningECC1, deeplearningECC3, deeplearningECC4, deeplearningECC5, deeplearningIntroduction}. It has been shown that, with short-to-medium length {(i.e., up to a few hundred bits)} codewords, DNN-based decoding can achieve competitive bit error rate (BER) performance. However, since the number of candidate codewords becomes extremely large with medium-to-large codeword lengths {(i.e., a few hundred to a few thousand bits)}, direct application of DNNs to ECC decoding becomes difficult because of the explosive number of layers and weights. In \cite{deeplearningECC4}, DNNs were employed on sub-blocks of the decoder, which were then connected via belief propagation decoding to enable scaling of deep learning-based ECC decoding. In \cite{deeplearningECC5}, the authors proposed recurrent neural network (RNN)-based decoding for linear codes, which outperforms the standard belief propagation (BP) decoding and significantly reduces the number of parameters compared to BP feed-forward neural networks.

To the best of our knowledge, no other work has yet been reported that explores deep learning-based decoding for CS codes. As we will show in the rest of our paper, deep learning fits well with CS decoding for both {FL} CS codes and {VL} CS codes, which we discuss separately in this paper. For {FL} codes, the deep learning-based decoder outperforms traditional {LUT} decoding, while naturally avoiding the explosive number of layers and weights that occur in ECC decoding. For {VL} codes, the deep learning-based decoder enables one-shot batch-processing of received sequences such that system throughput is improved, while simultaneously providing stronger error-correction capability.

Throughout this paper we focus on two types of CS codes for wireless communications: DC-free codes that have been employed in visible light communications \cite{VLC1}, and runlength-limited (RLL) codes that have been proposed to realize efficient wireless energy harvesting \cite{energyHarvesting,energyHarvestingRLL1}. However, we stress that our discussion applies to any other CS codes.

The contributions of this paper are as follows. For {FL} CS codes:
\begin{itemize}
	\item We explore multiple layer perception (MLP) networks and convolutional neural networks (CNNs) for {FL} CS decoding, and show that use of a CNN {reduces the number of parameters that need to be trained} by employing the constraints that are inherent in CS codewords.
	\item We show that well-trained networks achieve BER performance that is very close to maximum a posteriori probability (MAP) decoding of {FL} CS codes, therefore increasing the reliability of transmission.
	
	\item {We demonstrate that the implementation of FL capacity-achieving CS codes with long codewords, which has been considered impractical, becomes practical with deep learning-based CS decoding.}
\end{itemize}

For {VL} CS codes:
\begin{itemize}
	\item We show that for both single-state {VL} and multi-state {VL} codes, CNNs are able to perform segmentation of codewords in the received sequences in one shot, therefore enabling batch-processing of received sequences by the {VL} CS decoder, which increases the system throughput.
	\item We demonstrate that with erroneous received sequences, a well-designed CNN exhibits error-correction capabilities such that it might still be able to segment erroneous sequences into codewords correctly and therefore maintain synchronization when it is difficult for traditional bit-by-bit processing to achieve similar performance.
\end{itemize}

We first provide background information before considering {FL} and {VL} codes in turn.

\section{Preliminaries}
\subsection{CS codes}\label{CS_codes}

CS encoders convert source bits into coded sequences that satisfy certain constraints imposed by the physical channel. Some of the most widely-recognized constraints include $(d,k)$ RLL constraints that bound the number of logic zeros between consecutive logic ones to be between $d$ and $k$, and DC-free constraints that bound the running digital sum (RDS) value of the encoded sequence, where RDS is the accumulation of encoded bit weights in a sequence given that a logic one has weight $+1$ and a logic zero has weight $-1$ \cite{Textbook}. Some other types of constraints include the Pearson constraint and constraints that mitigate inter-cell interference in flash memories \cite{Pearson1, Pearson2, SystematicPearsonCoding,Motwani,MyMinimalSet,MyJSAC,myPearson,My_MinimalSet}.

\begin{figure*}[htbp]
	\begin{center}
		\includegraphics[width=12.5cm]{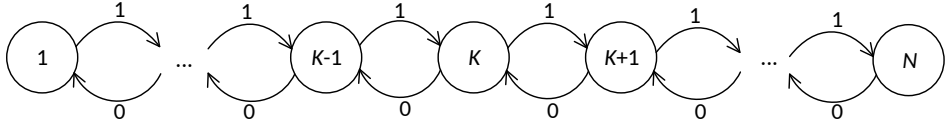}
		\makeatletter\def\@captype{figure}\makeatother
		\caption{FSM of DC-free constraints}\label{general}
	\end{center}
\end{figure*}

CS encoders can be described by {finite state machines (FSMs)} consisting of states, edges and labels. For example, the FSM of a DC-free constraint with $N$ different RDS values is shown in Fig. \ref{general}. The \emph{capacity} of a constrained sequence $C$ is defined as \cite{Shannon}
\begin{equation}\label{eq1}
C=\lim_{m \rightarrow \infty}\frac{\log_2 {\cal N}(m)}{m}
\end{equation}
where ${\cal N}(m)$ denotes the number of constraint-satisfying sequences of length $m$. Based on the FSM description and the adjacency matrix $\bf{D}$ \cite{Textbook}, we can evaluate the capacity of a constraint by calculating the logarithm of $\lambda_{max}$ which is the largest real root of the determinant equation \cite{Shannon}
\begin{equation}\label{eq2}
\det[{\bf{D}} - z{\bf{I}}] = {\bf{0}}
\end{equation}
where $\bf{I}$ is an identity matrix. The capacity is given as \cite{Shannon}

\begin{equation}\label{eq3}
C=\log_2\lambda_{max}
\end{equation}
with units bits of information per symbol.

\begin{figure*}[htbp]
	\begin{center}
		\includegraphics[width=130mm]{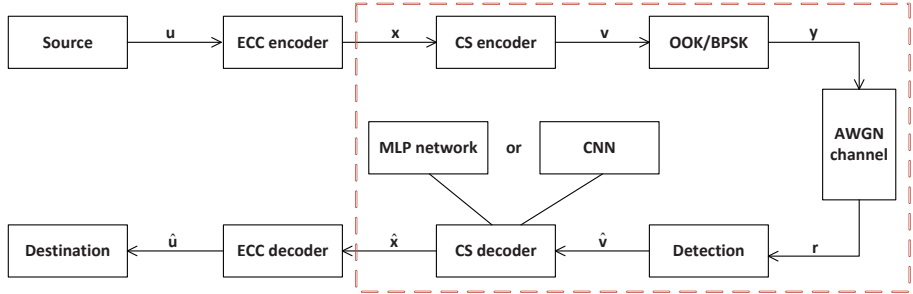}
		\makeatletter\def\@captype{figure}\makeatother
		\caption{System model}\label{system_model}
	\end{center}
\end{figure*}

\subsection{System model}	

We consider a typical wireless communication system as shown in Fig. \ref{system_model}. In particular, we consider visible light communications when demonstrating the use of DC-free codes, and wireless energy harvesting when demonstrating the use of RLL codes. Source bits $\bf{u}$ are encoded by an ECC encoder and a CS encoder to generate coded bits $\bf{x}$ and $\bf{v}$, respectively. The coded bits are then modulated to $\bf{y}$ and transmitted via an additive white Gaussian noise (AWGN) channel. We use on-off keying (OOK) modulation for VLC, and binary phase shift keying (BPSK) for wireless energy harvesting. The received bits are
\begin{equation}
{\bf{r}} = {\bf{y}} + {\bf{n}}
\end{equation}
where ${\bf{n}}$ is the noise vector where each element is a Gaussian random variable with a zero mean and variance $\sigma^2$. The detector outputs symbol estimates $\bf{\hat{v}}$, and this sequence of estimates is decoded with the CS decoder and ECC decoder successively to generate the estimate $\bf{\hat{u}}$. In this paper we consider the framed components, and focus on the CS decoder that outputs $\bf{\hat{x}}$ as close as possible to $\bf{x}$. Throughout the paper we denote $|\bf{x}|$ as the size of vector $\bf{x}$.	

\subsection{MLP networks and CNNs}
The fundamentals of deep learning are comprehensively described in \cite{deeplearningBook}. We employ both MLP networks and CNNs for CS decoding to predict $\bf{\hat{x}}$ given the input $\bf{\hat{v}}$. An MLP network has $L$ feed-forward layers. For each of the neurons in the MLP network, the {output $\nu$} is determined by the input vector $\bf{t}$, the weights vector ${\bm{\theta}}$ and the activation function $g()$:
\begin{equation}
{\nu = g({\boldsymbol{\theta}}\bf{t})}
\end{equation}
where for the activation function we use the sigmoid function {$g(z) = \frac{1}{1 + \exp(-z)}$} and the rectified linear unit (ReLU) function {$g(z) = \max\{0,z\}$}. A deep MLP network consists of many layers; the $i$th layer performs the mapping ${\bf{f}}^{(i)}:\mathbb{R}^{t_i} \rightarrow \mathbb{R}^{m_i}$, where $t_i$ and $m_i$ are the lengths of the input vector and the output vector of that layer, respectively. The MLP network is represented by:
\begin{equation}
{\bf{\hat{x}}} = {\bf{f}}^{(L)}({\bf{f}}^{(L-1)}(...{\bf{f}}^{(2)}({\bf{f}}^{(1)}(\bf{\hat{v}}))))
\end{equation}

The use of CNNs has recently achieved impressive performance in applications such as visual recognition, classification and segmentation, among others \cite{deeplearningBook}. It employs convolutional layers to explore local features instead of extracting global features with fully connected layers as in MLP networks, thus greatly reducing the number of weights that need to be trained and making it possible for the network to grow deeper \cite{resnet}. Different from visual tasks where the input colored images are represented by three dimensional vectors, the input vector $\bf{\hat{v}}$ in our task of CS decoding is a one dimensional vector. For a convolutional layer with $F$ \emph{kernels} given as: ${\bf{q}}^{f} \in \mathbb{R}^{1 \times |{\bf{q}}|}, {f = 1,...,F}$, the generated \emph{feature map} ${\bf{\hat{p}}}^f \in \mathbb{R}^{1 \times |{\bf{\hat{p}}}|}$ from the input vector ${\bf{\hat{v}}} \in \mathbb{R}^{1 \times |{\bf{\hat{v}}}|}$ satisfies the following dot product:
\begin{equation}
{p}^{f}_i = \sum_{l=0}^{|{\bf{q}}|-1}q^f_l {\hat{v}}_{1+s(i-1)+l}
\end{equation}
where $s \ge 1$ is the \emph{stride}. {CNNs benefit from weight sharing that exploits the spatial correlations of images and thus usually require fewer weights to be trained.} Usually convolutional layers are followed by \emph{pooling layers} such that high-level features can be extracted at the top layers. However, as we will show, pooling may not fit well in CS decoding and therefore our implementation of CNN does not include pooling layers.

\section{Deep learning-based decoding for fixed-length CS codes}\label{fixedlength}
\subsection{The fixed-length 4B6B code in visible light communications}

VLC refers to short-range optical wireless communication using the visible light spectrum from 380 nm to 780 nm, and has gained much attention recently \cite{VLC1}. The simplest VLC relies on OOK modulation, which is realized with DC-free codes to generate a constant dimming level of $50\%$. Three types of {FL} DC-free codes have been used in VLC standards to adjust dimming control and reduce flicker perception: the Manchester code, the 4B6B code and the 8B10B code \cite{VLC1}. We use the 4B6B code as a running example in this section to discuss {FL} CS codes.

The 4B6B code satisfies the DC-free constraint with $N=5$, which has a capacity of 0.7925 \cite{Textbook}. The codebook has 16 source words as shown in Table \ref{4B6B} \cite{VLC1}. Each source word has a length of 4 and is mapped to a codeword of length 6, which results in a code rate $R$ of $2/3$, and therefore an efficiency $\eta = R/C$ of $84.12\%$. In the remainder of this section we discuss how we use DNNs to decode the 4B6B code. We employ both MLPs and CNNs, and compare their performance.

\begin{table}[htbp]
	\centering
	\caption{Codebook of the 4B6B DC-free VLC code: $R=2/3, \eta=84.12\% $}\label{4B6B}
	\begin{tabular}{|c|c||c|c|}
		\hline
		Source word&  Codeword & Source word&  Codeword   \\
		\hline
		0000    &    001110   &  1000    &     011001  \\
		\hline
		0001   &     001101   &  1001   &     011010  \\
		\hline
		0010    &     010011  &  1010& 011100  \\
		\hline
		0011    &     010110  &  1011    &    110001  \\
		\hline
		0100    &     010101 &  1100    &    110010 \\
		\hline
		0101    &     100011 &  1101    &    101001 \\
		\hline
		0110    &     100110 &  1110    &    101010 \\
		\hline
		0111    &     100101 &  1111    &    101100 \\
		\hline
	\end{tabular}
\end{table}

\subsection{Training method}

In order to constrain the size of the training set, we follow the training method in \cite{deeplearningECC1} where the DNN was extended with additional layers of modulation, noise addition and detection that have no additional parameters that need to be trained. Therefore, it is sufficient to work only with the sets of all possible noiseless codewords ${\bf{v}} \in \mathbb{F}_2^{|\bf{v}|}, \mathbb{F}_2 \in \{0,1\}$, i.e., training epoches, as input to the DNNs. For the additional layer of detection, we calculate the log-likelihood ratio (LLR) of each received bit and forward it to the DNN.
We use the mean squared error (MSE) as the loss function, which is defined as:
\begin{equation}\label{MSE}
L_{MSE} = \frac{1}{|{\bf{x}}|} \sum_i(x_i-\hat{x}_i)^2.
\end{equation}

Both the MLP networks and CNNs employ three hidden layers, details of which are discussed in the next section. We aim at training a network that is able to generalize, that is we train at a particular signal-to-noise ratio (SNR) and test it within a wide range of SNRs. The criterion for model selection that we employ follows \cite{deeplearningECC1}, which is the normalized validation error (NVE) defined as:
\begin{equation}\label{NVE}
NVE(\rho_t) = \frac{1}{S}\sum_{s=1}^S \frac{BER_{DNN}(\rho_t,\rho_{v,s})}{BER_{MAP}(\rho_{v,s})},
\end{equation}
where $\rho_{v,s}$ denotes the $S$ different test SNRs. $BER_{DNN}(\rho_t,\rho_{v,s})$ denotes the BER achieved by the DNN trained at SNR $\rho_t$ and tested at SNR $\rho_{v,s}$, and $BER_{MAP}(\rho_{v,s})$ denotes the BER of MAP decoding of CS codes at SNR $\rho_{v,s}$. The networks are trained with {a fixed number of epoches that we will present in the next section.}

\subsection{Results and outlook}
We use the notation {${\bf{h}} = [h_1, h_2,...,h_L]$} to represent a network with $L$ hidden layers, where $h_l$ denotes the number of neurons in the fully connected layer $l$, or the number of kernels in the convolutional layer $l$. In recent works that apply DNNs to decode ECCs, the training set explodes rapidly as the source word length grows. For example, with a rate 0.5 $(n=1024, k=512)$ ECC, one epoch consists of $2^{512}$ possibilities of codewords of length 1024, which results in very large complexity and makes it difficult to train and implement DNN-based decoding in practical systems \cite{deeplearningECC1, deeplearningECC2, deeplearningECC3, deeplearningECC4}. However, we note that in {FL} CS decoding, this problem does not exist since CS source words are typically considerably shorter, possibly only up to a few dozen symbols \cite{Textbook, Franaszek, 8B10B_Ref,DC-freeRLL_Immink,FixedLength2,ImminkVLCodes,Motwani, BalancedModulation,FixedLength3,Pearson1,SystematicPearsonCoding,Pearson2,myDeepLearning}. This property fits deep learning based-decoding well.
\subsubsection{BER performance}\label{BER}
\paragraph{Frame-by-frame decoding}\label{1frame}
First we consider frame-by-frame transmission, where the 4B6B codewords are transmitted and decoded one-by-one, i.e., $|{\bf{v}}|=6$. We will later consider processing multiple frames simultaneously to improve the system throughput. Note that in the VLC standard, two 4B6B look-up tables can be used simultaneously \cite{VLC1}.

We compare performance of deep learning-based decoding with conventional {LUT} decoding that generates hard-decision bits. That is, the traditional detector estimates the hard decision $\hat{\bf{v}}$, and the CS decoder attempts to map $\hat{\bf{v}}$ to a valid source word to generate $\hat{\bf{x}}$. If the decoder is not able to locate $\hat{\bf{v}}$ in the code table due to erroneous estimation at the detector, the decoder determines the codeword that is closest to $\hat{\bf{v}}$ in terms of Hamming distance, and then outputs the corresponding source word. We also implement the maximum likelihood (ML) decoding of CS codes, where the codeword with the closest Euclidean distance to the received noisy version of the codeword is selected and the corresponding source word is decoded. We assume equiprobable zeros and ones in source sequences, thus ML decoding is equivalent to MAP decoding of CS codes since each codeword has an equal occurrence probability.


\begin{table*}[htbp]
	\centering
	\caption{Parameters of the MLP networks and CNNs trained for CS decoding with different frames}\label{parameters}
	\begin{tabular}{|c||c|c||c|c||c|}
		\hline  & \multicolumn{2}{|c||}{MLP}  & \multicolumn{2}{|c||}{CNN} &\\
		\hline
		\# of frame&  \# of neurons & \# of parameters  &   \# of kernels & \# of parameters & epoches  \\
		\hline
		1    &  [32,16,8]    &   924   &  [8,12,8]    & 760 & 4$\mathrm{e}$+4\\
		\hline
		2   &     [64,32,16]    & 3576    &  [8,14,8]    & 1374 & 3$\mathrm{e}$+4\\
		\hline
		3    &     [128,64,32]  &  13164 &  [8,16,8] & 2372 & 3$\mathrm{e}$+4\\
		\hline
		4    &     [128,128,64]   &  29008 &     [16,16,12]  & 5676 & 2.5$\mathrm{e}$+3\\
		\hline
		5    &     [256,128,64]  &   50338 &    [16,32,12] & 9536 & 2.5$\mathrm{e}$+3\\
		\hline
	\end{tabular}
\end{table*}	

Table \ref{parameters} shows the parameters of the MLP networks and the CNNs for a variety of tasks, {and the number of epoches used for each network}. The DNNs are {initialized using \emph{Xavier initialization} \cite{Xavier}}, and trained at an SNR of 1 dB using \emph{Adam} for stochastic gradient descent optimization \cite{Adam}. With $|{\bf{v}}| = 6$, { the parameters of the proposed networks are shown in row three of Table \ref{parameters}.} The MLP network we trained for frame-by-frame decoding has three hidden layers [32,16,8] and 924 trainable parameters. Its BER performance is shown in Fig. \ref{1frame}, which shows that DNN-based decoding achieves a BER that is very close to MAP decoding of CS codes, and outperforms the conventional {LUT} decoding described above by $\sim$2.2 dB.

\begin{figure}[htbp]
	\begin{center}
		\includegraphics[width=0.9\linewidth]{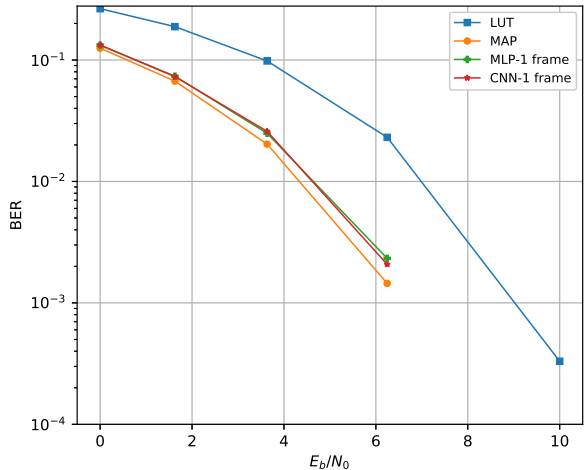}
		\makeatletter\def\@captype{figure}\makeatother
		\caption{Comparison of BER performance with frame-by-frame transmission, i.e., $|{\bf{v}}|=6$}\label{1frame}
	\end{center}
\end{figure}

We then investigate employing CNNs for this task. With ECC decoding, CNNs and MLP networks have similar number of weights in order to achieve similar performance \cite{deeplearningECC3}. In the following we outline our findings that are unique to CS decoding.

\begin{table*}[htbp]
	\centering
	\caption{Structures of the CNNs for CS decoding}\label{cnnparameters}
	\begin{tabular}{|c|c|c|c|}
		\hline
		layer  &  kernal size / stride & input size   & padding \\
		\hline
		OOK    &  N/A    &  $1 \times |{\bf{v}}|$  &  N/A\\
		\hline
		Adding noise   &   N/A      &  $1 \times |{\bf{v}}|$ &  N/A\\
		\hline
		LLR    &  N/A   & $1 \times |{\bf{v}}|$ &  N/A\\
		\hline
		Convolution   &   $1 \times 3$ / 1     & $1 \times |{\bf{v}}|$    &  no    \\
		\hline
		Convolution    &  $1 \times 3$ / 1    &  $1 \times (|{\bf{v}}|-2) \times h_1$    &  yes  \\
		\hline
		Convolution    &  $1 \times 3$ / 1    &   $1 \times (|{\bf{v}}|-2) \times h_2$   &  yes  \\
		\hline
		Fully connected    &   N/A    &  $1 \times ((|{\bf{v}}|-2) \times h_3)$      &  N/A\\
		\hline
		Sigmoid    &  N/A     &  $1 \times (|{\bf{x}}|)$    &  N/A  \\
		\hline
	\end{tabular}
\end{table*}

In Table \ref{cnnparameters} we outline the structure of the CNNs we apply for CS decoding. ReLU is used as the activation function for each convolutional layer. We note that CS codes always have inherent constraints on their codewords such that they match the characteristics of the channel. For example, the 4B6B code in Table \ref{4B6B} always has an equal number of logic ones and logic zeros in each codeword, and the runlength is limited to four in the coded sequence for flicker reduction. These \emph{low-level features} can be extracted to enable CNNs to efficiently learn the weights of the kernels, which results in {a smaller number of weights that need to be trained in the training phase} compared to MLP networks. For example, although similar BER performance is achieved by the [32, 16, 8] MLP network and the [6, 10, 6] CNN, the number of weights in the CNN is only $82\%$ of that in the MLP network. With larger networks {the reduction in the number of weights that need to be trained is more significant,} as we show in the next subsection.

Another finding we observe during training of a CNN is that pooling layers, which are essential component structures in CNNs for visual tasks, may not be required in our task. The reason is that in visual tasks, pooling is often used to extract \emph{high-level} features of images such as shapes, edges or corners. However, CS codes often possess low-level features only, and we find that adding pooling layers may not assist CS decoding. Therefore, no pooling layer is used in our CNNs, as indicated in Table \ref{cnnparameters}. Fig. \ref{1frame} shows that use of a CNN achieves similar performance to the use of an MLP network, and that it also approaches the performance of MAP decoding.

\begin{table*}[htbp]
	\centering
	\caption{{Test phase time and space complexity of the MLP networks and CNNs used in Table \ref{parameters}}}\label{complexity1}
	\begin{tabular}{|c||c|c||c|c|}
		\hline  & \multicolumn{2}{|c||}{MLP}  & \multicolumn{2}{|c|}{CNN}\\			
		\hline
		\# of frame&  Time (FLOPs) & Space (MBytes) &  Time (FLOPs) & Space (MBytes)   \\
		\hline
		1    &  8.64$\mathrm{e}$+2   &   0.0035   &  2.53$\mathrm{e}$+3    & 0.0032\\
		\hline
		2   &     3.46$\mathrm{e}$+3    & 0.014    & 7.6$\mathrm{e}$+3 & 0.0063\\
		\hline
		3    &     1.29$\mathrm{e}$+4  &  0.05   & 1.42$\mathrm{e}$+4 & 0.011\\
		\hline
		4    &     2.87$\mathrm{e}$+4   &  0.11    &  3.48$\mathrm{e}$+4   & 0.025\\
		\hline
		5    &     4.99$\mathrm{e}$+4 &  0.19 &  8.33$\mathrm{e}$+4 & 0.043\\
		\hline
	\end{tabular}
\end{table*}

\paragraph{Improving the throughput}\label{nframe}
We now consider processing multiple frames in one time slot in order to improve the system throughput. The system throughput can be enhanced by increasing the optical clock rate, which has its own physical limitations, or by processing multiple 4B6B codewords in parallel. The VLC standard allows two 4B6B codes to be processed simultaneously \cite{VLC1}. Now we show that DNNs can handle larger input size where $|{\bf{v}}|$ is a multiple of 6, thus system throughput can be enhanced by using one of those DNNs or even using multiple DNNs in parallel.

\begin{figure}[htbp]
	\begin{center}
		\includegraphics[width=0.9\linewidth]{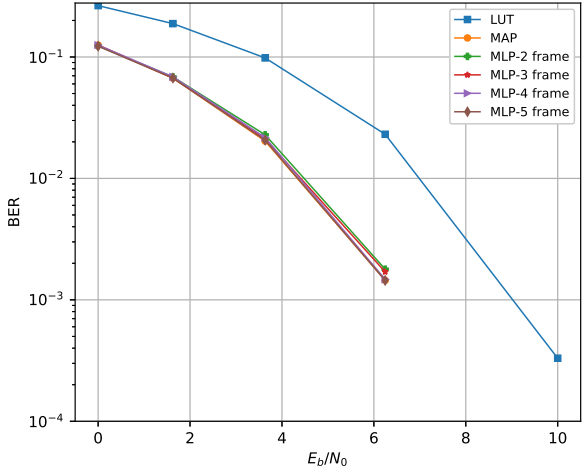}
		\makeatletter\def\@captype{figure}\makeatother
		\caption{BER performance of MLP networks with multiple frames processed simultaneously, $|{\bf{v}}|=12, 18, 24, 30$}\label{nframe_MLP}
	\end{center}
\end{figure}

\begin{figure}[htbp]
	\begin{center}
		\includegraphics[width=0.9\linewidth]{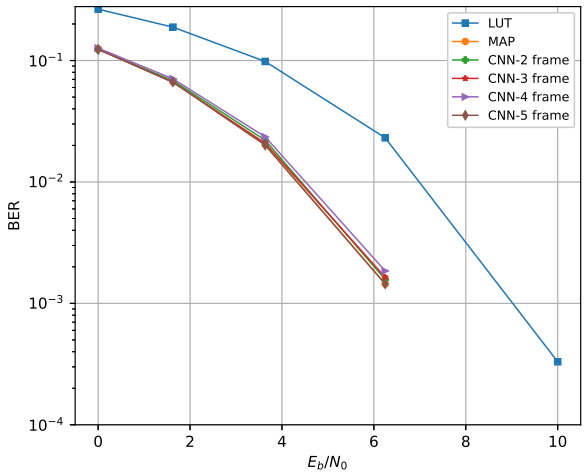}
		\makeatletter\def\@captype{figure}\makeatother
		\caption{BER performance of CNNs with multiple frames processed simultaneously, $|{\bf{v}}|=12, 18, 24, 30$}\label{nframe_CNN}
	\end{center}
\end{figure}

Figs. \ref{nframe_MLP} and \ref{nframe_CNN} present the BER performance of MLP networks and CNNs respectively. The parameters of those networks are shown in Table \ref{parameters}, { where rows 3, 4, 5 and 6 correspond to the parameters of the proposed networks with $|{\bf{v}}|=12,18,24,30$, respectively.} These figures demonstrate that both MLP networks and CNNs are able to achieve BERs very close to MAP decoding, while the CNNs have {a smaller number of weights that need to be trained at the training phase} compared to the MLP networks for the reasons outlined above. With larger $|{\bf{v}}|$, CNN becomes more advantageous in extracting the low-level features from the longer input sequences and learning the weights, and thus the reduction in the number of trainable parameters with a CNN is more significant. For example, when processing five frames simultaneously, the CNN has less than $1/5$ of the parameters that need to be trained for the MLP network. Note that we consider the number of parameters in the networks shown in Table \ref{parameters} to be small, e.g., ResNet \cite{resnet} has a few million parameters to train. 

{ Following the study in \cite{CVPR1,Complexity}, we now consider the time and space complexities of the proposed networks at the test phase. We study the theoretical time complexity instead of actual running time, because the actual running time can be sensitive to specific hardware and implementations. The time complexity of an MLP network is measured, with units of floating-point operations (FLOPs), as:
	\begin{equation}\label{timecomplexityMLP}
	O \bigg( \sum_{l=1}^{L+1} h_{l-1}  \cdot h_{l} \bigg),
	\end{equation}	
	where $h_0=|\bf{v}|$ and $h_{L+1}=|\bf{x}|$.
	
	The space complexity of the MLP networks is measured, with units of bytes, as:
	\begin{equation}\label{spacecomplexityMLP}
	O \bigg( 4\sum_{l=1}^{L+1} \Big( h_{l-1} \cdot h_{l} + (1 \cdot m_l) \Big) +4|\bf{v}|\bigg),
	\end{equation}
	where $m_l$ is the length of the output of layer $l$ and $m_{L+1}=|\bf{x}|$. $h_{l-1} \cdot h_{l}$ is the memory needed to store the weights of the network, $1 \cdot m_l$ is the memory needed to store the output of each layer, and $|\bf{v}|$ is the memory needed to store the input vector.
	
	For CNNs, the time complexity of a CNN is measured, with units of floating-point operations (FLOPs), as:
	\begin{equation}\label{timecomplexityCNN}
	O \bigg( \sum_{l=1}^{L+1} h_{l-1} \cdot (1 \cdot {\cal K}_l) \cdot h_{l} \cdot (1 \cdot m_l) \bigg),
	\end{equation}
	where $h_0=1$ and $h_{L+1}=|\bf{x}|$, and ${\cal K}_l$ denotes the length of the kernel at layer $l$. $m_l$ is the length of the output of layer $l$. ${\cal K}_{L+1}=m_L$, $m_{L+1}=1$ \cite{LeNet}.
	
	The space complexity of the CNNs is measured, with units of bytes, as:	
	\begin{equation}\label{spacecomplexityCNN}
	O \bigg( 4\sum_{l=1}^{L+1} \Big(  (1 \cdot {\cal K}_l) \cdot h_{l-1} \cdot h_{l} + (1 \cdot m_l) \cdot h_l \Big) +4|\bf{v}| \bigg).
	\end{equation}

	In Table \ref{complexity1}, we provide the time and space complexity of the networks we described in Table \ref{parameters}. It is seen that, at the test phase, CNNs have higher time complexity since convolutional layers are time-consuming at the test phase \cite{CVPR1}. However, CNNs have less space complexity due to weight sharing, and they require less memory. Nevertheless, we note that all networks listed in Table \ref{parameters} are small and have relatively low implementation complexity when compared to other modern DNNs, for example, VGG-16 \cite{VGG16} which has time complexity of 1.55$\mathrm{e}$+10 FLOPs and space complexity of 5.53$\mathrm{e}$+2 MB \cite{CVPR1}. Therefore, we anticipate that it will be practical to achieve further improvement of system throughput with larger and deeper networks.

}

\subsubsection{Paving the way to {FL} capacity-achieving CS codes}\label{capacity-achieving}

As we outlined in Section I, it is not an easy task to implement capacity-achieving {FL} CS codes. As determined by equations \eqref{eq1}-\eqref{eq3}, the capacity of a constraint is irrational in all but a very limited number of cases \cite{irrationalCapacity}, and can be typically approached with fixed-length codes of rate $R=k/n$ only with very large $k$ and $n$ values. This hinders implementation of {LUT} encoding and decoding for capacity-achieving CS codes. For example, as shown in Section II-B, the 4B6B code achieves $84.12\%$ of the capacity of a DC-free code with 5 different RDS values, which has $C=0.7925$. In Table \ref{knvalues} we list, for increasing values of $k$, values of $n$ that result in high code rates. This table shows that with $k=11,15,19$, it could be possible to construct fixed-length codes with efficiencies that exceed $99\%$. With $k$ as large as $79$ and $n=100$, the code would have rate 0.79 and efficiency $99.68\%$. However, an {LUT} codebook with $2^{k}$ source word-to-codeword mappings becomes impractical to implement as $k$ grows large. Other examples are the $k$-constrained codes recently developed for DNA-based storage systems in \cite{DNAStorageImmink}. Those fixed-length 4-ary $k$-constrained codes have rates very close to the capacity, however they require very large codebooks. For example, with method B in \cite{DNAStorageImmink}, for the $k$-constrained code with $k=1,2,3,4$, the codebooks have $4^{10},4^{38},4^{147},4^{580}$ source word-to-codeword mappings respectively.

\begin{table}
	\centering
	\caption{Highest code rate and efficiency with fixed-length CS codes for the DC-free constraint with five different RDS values, $C=0.7925$}\label{knvalues}
	\begin{tabular}{|c|c|c|c||c|c|c|c|}
		\hline
		$k$&$n$&$R$&$\eta$&$k$&$n$&$R$&$\eta$ \\
		\hline
		1&2&0.5000&63.09\%&11&14&0.7857&99.14\%\\
		2&3&0.6667&84.12\%&12&16&0.7500&94.64\%\\
		3&4&0.7500&94.64\%&13&17&0.7647&96.49\%\\
		4&6&0.6667&84.12\%&14&18&0.7778&98.14\%\\
		5&7&0.7143&90.13\%&15&19&0.7895&99.62\%\\
		6&8&0.7500&94.64\%&16&21&0.7619&96.14\%\\
		7&9&0.7778&98.14\%&17&22&0.7727&97.51\%\\
		8&11&0.7273&91.77\%&18&23&0.7826&98.75\%\\
		9&12&0.7500&94.64\%&19&24&0.7917&99.89\%\\
		10&13&0.7692&97.06\%&20&26&0.7692&97.06\%\\
		\hline
	\end{tabular}
\end{table}

With DNNs, however, it becomes practical to handle a large set of source word-to-codeword mappings which {used to be} considered impractical with {LUT} decoding. This paves the way for practical design and implementation of fixed-length capacity-achieving CS codes. Appropriate design of such codes is a practice of using standard algorithms from the rich theory of CS coding, such as Franaszek's recursive elimination algorithm \cite{Franaszek}, or the sliding-block algorithm \cite{ACH1, ACH2} with large $k$ and $n$ values to determine the codebooks. We propose implementing both the encoder and the decoder with DNNs. Although here we focus on decoding, similar to DNN-based decoders, CS encoders map noiseless source words to codewords, and can also be implemented with DNNs.

\begin{figure}[htbp]
	\begin{center}
		\includegraphics[width=0.9\linewidth]{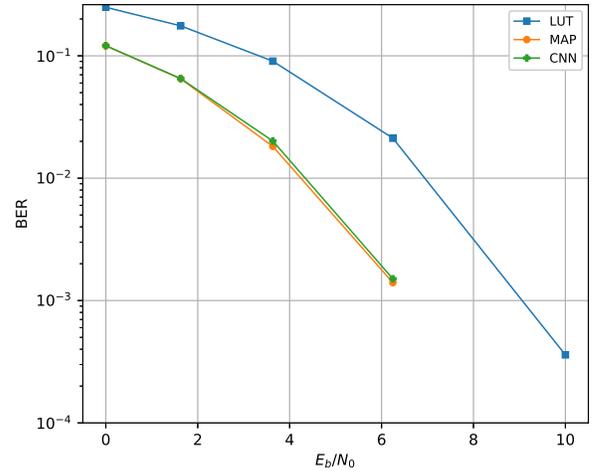}
		\makeatletter\def\@captype{figure}\makeatother
		\caption{BER performance of DNN-based decoding for a constrained sequence code with $2^{20}$ possibilities of mappings }\label{CNN_shuffle}
	\end{center}
\end{figure}

We now demonstrate that the proposed DNN-based \;decoders are able to map long received words from a CS code to their corresponding source words. We concatenate five 4B6B codebooks, where each 4B6B codebook is randomly shuffled in terms of its source word-to-codeword mappings, to generate a large codebook with $2^{20}$ entries of mappings. We train and test a CNN with ${\bf{h}} = [16,32,12]$ that has 9536 weights to perform decoding. From Fig. \ref{CNN_shuffle}, we can see that this CNN is capable of decoding the received noisy version of the large set of received words. The BER is close to MAP decoding, and outperforms the {LUT} decoding approach that we implemented as a benchmark. Therefore, the design and implementation of DNN-based CS decoding is practical with long CS codewords because of their low-level features that we can use to our advantage to simplify decoding.

We also note that DNNs that are proposed in communication systems could have many more parameters than the 9536 we use in this network. For example, \cite{SCMA} proposes an MLP network with 4 hidden layers where each layer has 512 neurons, which results in at least $512\times512\times3=786432$ parameters. Thus a larger CNN can be trained to decode fixed-length capacity-approaching CS codes with longer codewords. Having found that CNNs require many fewer parameters to train compared to MLP networks, we continue using CNNs for decoding variable-length CS codes in the next section.

\section{Deep learning-based decoding for {VL} CS codes}\label{variablelength}
\subsection{{VL} CS codes}

It has been recently reported in the literature that {VL} CS codes can exhibit higher code rates and simpler codebooks than their {FL} counterparts. Capacity-approaching {VL} codes have been designed with a single state \cite{MyMinimalSet,My_MinimalSet,MyJSAC,myPearson} and multiple states \cite{MyMultiState} in their codebooks. With a one-to-one correspondence between the lengths of {VL} source words and {VL} codewords, and the assumption of independent and equiprobable zeros and ones in the source sequence, the average code rate $\overline R$ is \cite{MyMinimalSet,My_MinimalSet}
\begin{equation}
\overline R  = \frac{{\bar k}}{{\bar n}}= \frac{{\sum\limits_{{s_i}}^{} {{2^{ - {s_i}}}} {s_i}}}{{\sum\limits_{{o_i}}^{} {{2^{ - {s_i}}}} {o_i}}}
\end{equation}
where $s_i$ is the length of the $i$-th source word that is mapped to the $i$-th codeword of length $o_i$. The efficiency of a code is defined as $\eta=\overline R/C$.

\subsubsection{Codes for wireless energy harvesting}

Wireless energy harvesting requires avoidance of battery overflows or underflows at the receiver, where overflow represents the event that energy is received but the battery is full, while underflow occurs when energy is required by the receiver when the battery is empty \cite{energyHarvesting}. Consider an energy harvesting system where logic one coded bits carry energy, bit logic zeros do not. In such cases, $(d,k)$ RLL codes have been proposed to work in regimes where overflow protection is the most critical. By substituting all logic zeros with logic ones and vice versa, use of these RLL codes have been proposed for regimes where underflow protection is the most critical \cite{energyHarvesting, energyHarvestingRLL1}.

Following the capacity-approaching VL code construction technique in \cite{MyMinimalSet,My_MinimalSet}, we construct a single-state codebook for the {VL} $(d=1,k=3)$ RLL code given in Table \ref{1_3dkCodebook} that achieves $98.9\%$ of capacity, and we use this code as a running example throughout this section. As a comparison, note that a widely
used $(d=1,k=3)$ {FL} RLL code is the MFM code with $\eta=91\%$ \cite{Textbook}, which demonstrates the increase in efficiency that is possible with use of {VL} CS codes.

\begin{table}[htbp]
	\centering
	\scriptsize
	\caption{Codebook of a $(d=1, k=3)$ RLL code with efficiency of $98.9\%$}\label{1_3dkCodebook}
	\begin{tabular}{|c|c|}
		\hline
		Source word & Codeword  \\
		\hline
		$0$       &   01     \\
		\hline
		$10$       &   001   \\
		\hline
		$11$       &   0001   \\
		\hline
	\end{tabular}
\end{table}

\subsubsection{Codes for visible light communications}
Table \ref{CodeBook_DCfree_N=5} (from \cite{MyMultiState}) presents a {VL} DC-free codebook that has two encoding states $\sigma_1, \sigma_2$ and satisfies exactly the same constraint as the 4B6B code in Table \ref{4B6B}. $\alpha(\sigma_i)$ and $\beta(\sigma_i)$ denote the output codewords and the next states. This code has an average code rate $\overline R = 0.7857$ and an average efficiency $\eta=99.14\%$, which is significantly higher than the 4B6B code. It also has fewer codewords compared to the 4B6B code.

\begin{table}[htbp]
	\centering
	\scriptsize
	\caption{Codebook of DC-free code with $\eta=99.14\%$ and an RDS value of 5}\label{CodeBook_DCfree_N=5}
	\begin{tabular}{|c|c|c|c|c|}
		\hline
		Source word &$\alpha(\sigma_1)$  & $\beta(\sigma_1)$ & $\alpha(\sigma_2)$ & $\beta(\sigma_2)$  \\
		\hline
		$00$ & $11$       &   $\sigma_2$   &  00  & $\sigma_1$     \\
		\hline
		$010$ & $0111$      &   $\sigma_2$   & 1000 &  $\sigma_1$     \\
		\hline
		$011$ & $0101$     &   $\sigma_1$   &  0101 &  $\sigma_2$     \\
		\hline
		$100$ & $0110$     &   $\sigma_1$   &  0110 &  $\sigma_2$     \\
		\hline
		$101$ & $1011$     &   $\sigma_2$   &  0100 &  $\sigma_1$     \\
		\hline
		$110$ & $1001$     &   $\sigma_1$   &  1001 &  $\sigma_2$     \\
		\hline
		$111$ & $1010$     &   $\sigma_1$   &  1010 &  $\sigma_2$     \\
		\hline
	\end{tabular}
\end{table}

In the following we demonstrate how we employ CNNs for {VL} CS decoders to improve the system throughput and reduce error rates.

\subsection{Conventional bit-by-bit processing}
Since the {VL} CS codes in Tables \ref{1_3dkCodebook} and \ref{CodeBook_DCfree_N=5} are instantaneous codes, when no errors occur during transmission, the received sequences can be accurately segmented at the decoder and decoded with bit-by-bit processing. Details of this process are given in Algorithm 1, where ${\bf{\widehat{v}}}_i^j$ denotes the sub-vector of ${\bf{\widehat{v}}}$ starting from the $i$th position and ending at the $j$th position. Whenever the receiver receives another bit, it attempts to match the current processed sequence with a codeword in the codebook. If the receiver is unable to find a match, it then takes the next received bit and repeats the process. For example, with the single-state $(d=1, k=3)$ RLL codebook in Table \ref{1_3dkCodebook}, the RLL coded sequence ${\bf{\widehat{v}}}=010001001$ is correctly decoded into $01110$ by the decoder that uses Algorithm 1.

\begin{algorithm}
	\caption{Conventional bit-by-bit decoding of error-free {VL} CS coded sequences}
	\begin{algorithmic}[1]
		
		\REQUIRE
		\STATE $\bf{\hat{v}}$, the codebook
		\STATE $cur\_pos\_head\gets1, cur\_pos\gets1$
		
		\ENSURE
		\WHILE {$cur\_pos\_head \le |\bf{\hat{v}}|$}
		\IF {${\bf{\hat{v}}}_{cur\_pos\_head}^{cur\_pos}$ is a valid codeword}
		\STATE decode ${\bf{\hat{v}}}_{cur\_pos\_head}^{cur\_pos}$ into the corresponding source word
		\STATE $cur\_pos \gets cur\_pos+1$;
		\STATE $cur\_pos\_head \gets cur\_pos$;
		
		\ELSE
		\STATE $cur\_pos \gets cur\_pos+1$;
		\ENDIF
		\ENDWHILE
	\end{algorithmic}
\end{algorithm}

If errors occur during transmission, the decoder will have low probability of segmenting the received sequence correctly, and hence may lose synchronization. For example, assume the encoded $(d=1,k=3)$ RLL sequence is ${\bf{v}}=010001001$, but that noise causes the fifth bit to be detected in error such that the erroneously detected sequence is ${\bf{\widehat{v}}}=010011001$. After correctly decoding codeword $01$ into 0, the bit-by-bit decoder would determine that the next codeword is 001 and hence would perform incorrect codeword segmentation and therefore would output an incorrect decoded sequence. Note that due to incorrect segmentation, synchronization is lost in the sense that codeword boundaries are {incorrectly determined}. This loss of synchronization, {i.e., incorrect identification of the start and end positions of each codeword,} may result in error propagation. One possible algorithm for decoding {VL} CS sequences in case of transmission errors is described in Algorithm 2 in the Appendix, which helps recover synchronization, but is unable to correct detection errors.

Therefore, there are two drawbacks of conventional decoding for {VL} CS codes: i) since decoding is processed bit-by-bit, the system throughput is limited especially when errors occur during transmission which requires more processing to be done, and ii) since bit-by-bit decoding is in fact greedy and takes into account only the \emph{local information} by considering only the next few bits when trying to find a codeword match, it is very likely that the decoder generates erroneous output and/or loses synchronization. To tackle these drawbacks, we propose using CNNs to help the decoder perform batch-processing to increase the system throughput and reduce the error rate such that the CNN-aided decoder can still generate correct output sequences in the event of transmission errors.

\subsection{CNN-aided decoding of {VL} CS codes}

\begin{figure}[htbp]
	\begin{center}
		\includegraphics[width=\linewidth]{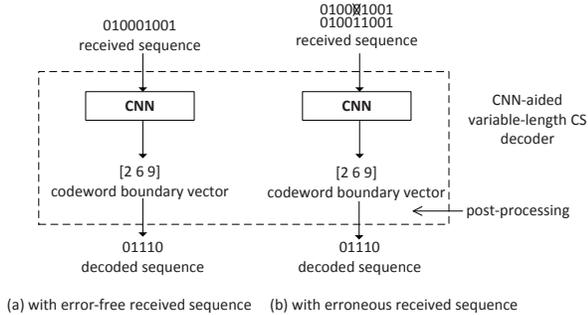}
		\makeatletter\def\@captype{figure}\makeatother
		\caption{Illustration of the CNN-aided {VL} CS decoder}\label{VLCNN}
	\end{center}
\end{figure}

We propose using CNNs for codeword segmentation of the received sequences to enable batch-processing. Consider the $(d=1,k=3)$ RLL code in Table \ref{1_3dkCodebook}. As illustrated in Fig. \ref{VLCNN}(a), $\bf{\widehat{v}}$ is passed into the decoder, where the CNN performs segmentation and generates a codeword boundary vector $\bf{\gamma}$ in one shot. Once $\bf{\gamma}$ is obtained, the decoder enters the post-processing phase and generates the source sequence by looking directly at the codeword boundaries and finding the corresponding source words in Table \ref{1_3dkCodebook}. In this way the system throughput can be improved. Moreover, as illustrated in Fig. \ref{VLCNN}(b), when determining the next codeword boundary, since the CNN considers an entire batch at a time, it also makes use of bits that are not close to the current codeword. In other words, the CNN not only makes use of \emph{local information}, but also takes advantages of \emph{global information} such that even if the received sequence is in error, the CNN may still be able to help the decoder stay synchronized and correctly recover the source sequence. Note that to take advantage of the full received information, in our implementation the input vector of the CNN is the noisy received sequence $\bf{r}$ instead of the demodulated bit sequence $\bf{\widehat{v}}$.

We propose that the codeword segmentation problem is analogous to two-dimensional object detection in computer vision. In object detection, the output is the coordinates $(a_1,b_1)$, $(a_1,b_2)$, $(a_2,b_1)$, $(a_2,b_2)$ of the four corners of the boxed rectangle that captures the detected object. Our CNN works with the one-dimensional input sequence $\bf{r}$, hence for the $i$th detected codeword in $\bf{r}$ it suffices to output two scalars $\gamma_i=(\gamma_{i,start},\gamma_{i,end})$ that correspond to the start and end positions of that codeword, where $\gamma_{i,end} = \gamma_{i+1,start} - 1$. Therefore, the output vector of the CNN is ${\boldsymbol  \gamma} = [\gamma_1, \gamma_2, ..., \gamma_{|{\bf{\gamma}}|}]$ which indicates the start and end positions of all codewords segmented from $\bf{r}$, which are exactly the synchronization positions. Furthermore, we propose to set the input size $|\bf{r}|$ of the CNN to be the largest packet size $L_{max}$ of the transmission system, such that the CNN is able to perform codeword segmentation with the largest packet and hence the start position of the first detected codeword is $\gamma_{1,start}=0$. Therefore, ${\boldsymbol  \gamma}$ is simplified to ${\boldsymbol  \gamma} = [\gamma_{1,end}, \gamma_{2,end}, ..., \gamma_{|{\bf{\gamma}}|,end}]$ where each element indicates a codeword boundary, which is a synchronization position. When the received packet has size smaller than $L_{max}$, the packet is augmented with padding symbols to match the input size of the CNN. Details of the padding are discussed below.

\paragraph{Training method}
The data generation phase is the same as in Section \ref{fixedlength}.B where the CNN is extended with additional layers that have no parameters such that it suffices to work only with the set of all possible noiseless codewords, except that we do not use the LLR layer in our CNN-aided {VL} CS decoder. From the codebook in Table \ref{1_3dkCodebook}, it is readily seen that the longest source sequence that can generate a coded sequence of length $L_{max}$ is $L_{max}/2$. We generate all possible source sequences of length $L_{max}/2$ and encode them into coded sequences that are at most of length $L_{max}$. If a received sequence has length smaller than $L_{max}$, it is padded with invalid symbol values {(which we denote here as --1)} such that the resulting length is $L_{max}$. The output vector $\boldsymbol \gamma$ has length $|\boldsymbol \gamma| = L_{max}/2$, where each of the elements is a codeword boundary, i.e., a synchronization position. If the number of codewords in the coded sequence is smaller than $L_{max}/2$, the rest of the elements are padded with symbols $L_{max}+1$ that indicate the last valid codeword boundary has been passed. At the post-processing phase, if the decoder encounters a value of $L_{max}+1$, it interprets this to indicate that the segmentation of the entire batch is completed, and it will finish decoding the current batch.

To illustrate, consider the situation in Fig. \ref{VLCNN}(a). If we set $L_{max}=10$, then the received sequence is padded with invalid symbol values, resulting in a vector of $[0,1,0,0,0,1,0,0,1,-1]$ that has length 10, and the output codeword boundary vector is ${\boldsymbol \gamma} = [2,6,9,11,11]$.

The criterion for model selection is similar to \eqref{NVE}, except that we employ the block error rates (BLERs) instead of the bit error rates. The NVE in this case is defined as:
\begin{equation}\label{NVEBLER}
NVE(\rho_t) = \frac{1}{S}\sum_{s=1}^S \frac{BLER_{CNN}(\rho_t,\rho_{v,s})}{BLER_{raw}(\rho_{v,s})},
\end{equation}
where $\rho_{v,s}$ denotes the $S$ different test SNRs. $BLER_{CNN}(\rho_t,\rho_{v,s})$ denotes the BLER of the decoded source sequence achieved by the CNN-aided decoder trained at SNR $\rho_t$ and tested at SNR $\rho_{v,s}$, and $BLER_{raw}(\rho_{v,s})$ denotes the raw BLER, i.e., the BLER of the received sequence $\bf{\widehat{v}}$, at SNR $\rho_{v,s}$. We use the MSE in \eqref{MSE} as the loss function. 

\begin{table*}[htbp]
	\centering
	\caption{Structures of the CNNs for {VL} CS decoding}\label{VLcnnparameters}
	\begin{tabular}{|c|c|c|c|}
		\hline
		layer  &  kernal size / stride & input size   & padding \\
		\hline
		BPSK/OOK    &  N/A    &  $1 \times |{\bf{r}}|$  &  N/A\\
		\hline
		Adding noise   &   N/A      &  $1 \times |{\bf{r}}|$ &  N/A\\
		\hline			
		Convolution   &   $1 \times 4$ / 1     & $1 \times |{\bf{r}}|$    &  no    \\
		\hline
		Convolution    &  $1 \times 5$ / 1    &  $1 \times (|{\bf{r}}|-3) \times h_1$    &  no  \\
		\hline
		Convolution    &  $1 \times 5$ / 1    &   $1 \times (|{\bf{r}}|-7) \times h_2$   &  no  \\
		\hline
		Fully connected    &   N/A    &  $1 \times ((|{\bf{r}}|-11) \times h_3)$      &  N/A\\
		\hline
		Fully connected    &   N/A    &  $1 \times 80$      &  N/A\\
		\hline
		Fully connected + ReLU    &   N/A    &  $1 \times 30$      &  N/A\\
		\hline
	\end{tabular}
\end{table*}

\paragraph{Results and outlook}

In our experiments we set ${|\bf{r}|}=L_{max}=12$. The structure of the CNN we developed is shown in Table \ref{VLcnnparameters}, where ${\bf{h}} = [16,32,20,80,30]$. The first three layers are convolutional layers and the last two layers are fully connected layers, resulting in 10188 trainable parameters. ReLU is used as the activation function for
each convolutional layer, and for the final output. {The number of epoches used for training this network is 1$\mathrm{e}$+5. We observed during experimentation that,} different from the prior-art (and Section \ref{fixedlength} in this paper) that employs deep learning for channel decoding where the DNNs are trained at relatively small values of SNR, in the codeword segmentation problem we find that the CNN should be trained at relatively large SNRs in order to work well. The intuitive reasoning for this observation is that { codeword segmentation is inspired from object detection in computer vision, and in object detection the CNN should be trained with ground-truth samples. Similarly, in the codeword segmentation problem, the CNN would be trained with codewords plus relatively small noise, i.e., ``ground-truth samples", as input.}

\begin{figure}[htbp]
	\begin{center}
		\includegraphics[width=0.9\linewidth]{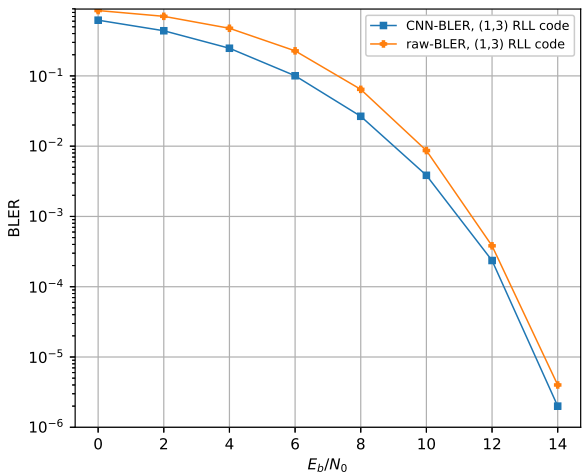}
		\makeatletter\def\@captype{figure}\makeatother
		\caption{Comparison of BLER performance for the {VL} $(d=1,k=3)$ RLL code}\label{VLCNNdk}
	\end{center}
\end{figure}

The result for the $(d=1,k=3)$ RLL code is shown in Fig. \ref{VLCNNdk}, where the CNN is trained at 10dB. It is clear that the CNN-aided decoder has smaller BLER than the raw BLER, indicating that the CNN is able to make use of global information to search for the global optimum instead of only the local optimum when performing codeword segmentation, and therefore has the capability of correctly detecting the codeword boundaries and maintaining synchronization even when the received sequence is in error. In low-to-medium SNR regions the CNN-aided decoder achieves $\sim$1 dB performance gain. Therefore, the CNN-aided decoder is not only able to improve the system throughput by enabling one-shot decoding, but it also has some error-correction capabilities.

\begin{figure}[htbp]
	\begin{center}
		\includegraphics[width=0.9\linewidth]{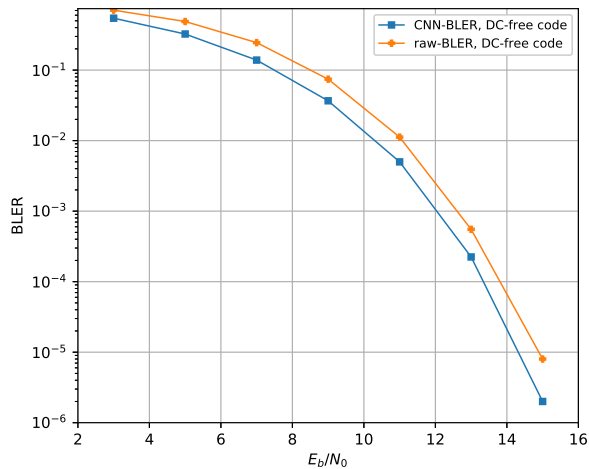}
		\makeatletter\def\@captype{figure}\makeatother
		\caption{Comparison of BER performance for the {VL} DC-free code}\label{VLCNNDCfree}
	\end{center}
\end{figure}

We now show that our approach also works for multi-state codes where state transitions occur during encoding and hence the state information is implicitly carried in coded sequences. We employ the multi-state DC-free code in Table \ref{CodeBook_DCfree_N=5} and train a CNN that has the same structure as in Table \ref{VLcnnparameters} to perform codeword segmentation. We note that, different from the $(d=1,k=3)$ RLL code in Table \ref{1_3dkCodebook}, although for this DC-free code the CNN is also able to perform codeword segmentation and help maintain synchronization, direct translation from $\bf{\gamma}$ to codewords is not possible since there is more than one codeword that has length 4. Therefore, during the post-processing phase the decoder performs soft-decision decoding based on the knowledge of $\bf{\gamma}$.
In Fig. \ref{VLCNNDCfree}, we present the BLER of source sequences after the CNN-aided decoder, and compare it to the raw BLER. The CNN is trained at 11dB. Similar to the $(d=1,k=3)$ RLL code, apart from the ability of increasing the system throughput, the CNN demonstrates error-correction capabilities such that it may still correctly perform codeword segmentation when errors occur during transmission, and also provides $\sim$1 dB performance gain in low-to-medium SNR regions. Therefore, the CNN-aided decoder is able to learn the inherent state information in the multi-state coded sequences, improve the system throughput, help maintain synchronization by correctly performing codeword segmentation, and reduce the error rate. Lastly, we note that those CNN-aided decoders are practical for {VL} CS codes, and that a larger batch could be processed with a larger CNN in an actual implementation.

\section{conclusion}
In this paper, we introduced deep learning-based decoding for {FL} and {VL} CS codes. For {FL} CS decoding, we studied two types of DNNs, namely MLP networks and CNNs, and found that both networks can achieve BER performance close to MAP decoding as well as improve the system throughput, where CNNs have {smaller number of trainable parameters} than MLP networks since they are able to efficiently exploit the inherent constraints imposed on the codewords. Furthermore, we showed that the design and implementation of {FL} capacity-achieving CS codes that {have been considered} impractical becomes practical with deep learning-based decoding, paving the way to deploying highly efficient CS codes in practical communication and data storage systems. We then developed CNNs for one-shot batch-processing of {VL} CS codes that exhibit higher code rates than their {FL} counterparts. We employed CNN-aided decoding for both single-state and multi-state {VL} CS codes, and showed that the CNN-aided decoder is not only able to improve the system throughput, but also has the capability of maintaining synchronization and performing error correction. 

\section*{Acknowledgment}

The authors are grateful to the anonymous reviewers for their comments that resulted in improvement of this paper.

\begin{appendices}
	\section{An algorithm for bit-by-bit decoding of {VL} CS codes}

	\begin{algorithm}
		\caption{Conventional bit-by-bit {VL} CS decoding when errors occur during transmission}
		\begin{algorithmic}[1]
			
			\REQUIRE
			\STATE $\bf{\hat{v}}$, the codebook
			\STATE $l_{max}$: the maximum length of codewords in the codebook
			\STATE $cur\_pos\_head \gets 1, cur\_pos \gets 1$
			
			\ENSURE
			\WHILE {$cur\_pos\_head \le |\bf{\hat{v}}|$}
			\IF {$cur\_pos - cur\_pos\_head + 1 \le l_{max}$}
			
			\IF {${\bf{\hat{v}}}_{cur\_pos\_head}^{cur\_pos}$ is a valid codeword}
			\STATE decode ${\bf{\hat{v}}}_{cur\_pos\_head}^{cur\_pos}$ into the corresponding source word
			\STATE $cur\_pos \gets cur\_pos+1$;
			\STATE $cur\_pos\_head \gets cur\_pos$;
			
			\ELSE
			\STATE $cur\_pos \gets cur\_pos+1$;
			\ENDIF
			
			\ELSE
			\STATE $/*$ \emph{no match is found in the codebook} $*/$
			\STATE $cur\_pos \gets cur\_pos\_head + 1$;
			
			\WHILE {$cur\_pos \le |\bf{\hat{v}}|$}
			\FOR{$tmp\_start\_pos$ in $[cur\_pos\_head, cur\_pos]$}
			\IF {${\bf{\hat{v}}}_{tmp\_start\_pos}^{cur\_pos}$ is a valid codeword}
			\STATE decode ${\bf{\hat{v}}}_{tmp\_start\_pos}^{cur\_pos}$
			\STATE $cur\_pos \gets cur\_pos+1$;
			\STATE $cur\_pos\_head \gets cur\_pos$;
			\STATE  goto line 5
			
			\ELSE
			\STATE $tmp\_start\_pos \gets tmp\_start\_pos + 1$
			\ENDIF
			\ENDFOR
			\STATE $cur\_pos \gets cur\_pos+1$;
			\ENDWHILE
			
			\ENDIF
			
			\ENDWHILE
		\end{algorithmic}
	\end{algorithm}
	
	One possible algorithm for bit-by-bit decoding of {VL} CS codes is shown in Algorithm 2. To illustrate, we continue our example in {Section \ref{variablelength}}.B where the transmitted sequence is $010001001$ and the detected sequence is $010011001$. According to this algorithm, after correctly decoding $01$ into 0 and incorrectly decoding $001$ into 10, the decoder processes lines 6--12 in Algorithm 2, tries to match $1,10,100,1001$ with codewords in Table \ref{1_3dkCodebook} sequentially and does not succeed. Since the decoder exhausted all possible codewords, it then processes lines 15--28, tries to match $1,10,0,100,00,0,1001,001$ with codewords in Table \ref{1_3dkCodebook} sequentially, finally finds that 001 is a match and decodes it into 10. The decoded sequence is $01010$. Note that although it does recover synchronization at the 9th received bit after loss of synchronization at the 5th received bit, the decoded sequence is incorrect due to temporary loss of synchronization, which contributes to the BLER.
	

\end{appendices}

\balance

\end{document}